\documentclass[aps,prapplied,superscriptaddress,notitlepage,twocolumn]{revtex4-1}

\usepackage[utf8]{inputenc}
\usepackage[english]{babel}
\usepackage{amsmath}
\usepackage{amsfonts}
\usepackage{amssymb}
\usepackage{graphicx}
\usepackage{hyperref}
\usepackage{pgf}
\usepackage{braket}
\usepackage{color,soul}

\DeclareUnicodeCharacter{2212}{$-$}

\newcommand{\Ibias}{\ensuremath{I}}

\newcommand{\Vmeas}{\ensuremath{V_\text{m}}}
\newcommand{\Vg}{\ensuremath{V_\text{G}}}

\newcommand{\Ic}{\ensuremath{I_\text{c}}}
\newcommand{\Icmax}{\ensuremath{I_{\text{c}\textrm{, max}}}}
\newcommand{\Rw}{\ensuremath{R_\text{S}}}

\newcommand{\VRF}{\ensuremath{V_\text{RF}}}

\let\pgfimageWithoutPath\pgfimage 
\renewcommand{\pgfimage}[2][]{\pgfimageWithoutPath[#1]{Figures/#2}}

\begin{document}

\title{Radio-frequency characterization of a supercurrent transistor made from a carbon nanotube}

\author{M. Mergenthaler}
\email[Now at IBM Quantum, IBM Research - Zurich, 8803 Rueschlikon, Switzerland: ]{mme@zurich.ibm.com}
\affiliation{Department of Materials, University of Oxford, Oxford OX1 3PH, United Kingdom}
\affiliation{Clarendon Laboratory, Department of Physics, University of Oxford, Oxford OX1 3PU, United Kingdom}

\author{F.J. Schupp}
\affiliation{Department of Materials, University of Oxford, Oxford OX1 3PH, United Kingdom}

\author{A. Nersisyan}
\affiliation{Clarendon Laboratory, Department of Physics, University of Oxford, Oxford OX1 3PU, United Kingdom}

\author{N. Ares}
\affiliation{Department of Materials, University of Oxford, Oxford OX1 3PH, United Kingdom}

\author{A. Baumgartner}
\affiliation{Department of Physics, University of Basel, Klingelbergstrasse 82, CH-4056 Basel, Switzerland}

\author{C. Sch\"onenberger}
\affiliation{Department of Physics, University of Basel, Klingelbergstrasse 82, CH-4056 Basel, Switzerland}

\author{G.A.D. Briggs}
\affiliation{Department of Materials, University of Oxford, Oxford OX1 3PH, United Kingdom}

\author{P.J. Leek}
\affiliation{Clarendon Laboratory, Department of Physics, University of Oxford, Oxford OX1 3PU, United Kingdom}

\author{E.A. Laird}
\email{e.a.laird@lancaster.ac.uk}
\affiliation{Department of Physics, Lancaster University, Lancaster LA1 4YB, United Kingdom}
\affiliation{Department of Materials, University of Oxford, Oxford OX1 3PH, United Kingdom}
	
\date{\today}

\begin{abstract}
A supercurrent transistor is a superconductor-semiconductor hybrid device in which the Josephson supercurrent is switched on and off using a gate voltage.
While such devices have been studied using DC transport, radio-frequency measurements allow for more sensitive and faster experiments.
Here a supercurrent transistor made from a carbon nanotube is measured simultaneously via DC conductance and radio-frequency reflectometry.
The radio-frequency measurement resolves all the main features of the conductance data across a wide range of bias and gate voltage, and many of these features are seen more clearly.
These results are promising for measuring other kinds of hybrid superconducting devices, in particular for detecting the reactive component of the impedance, which a DC measurement can never detect.
\end{abstract}
\maketitle

\section{Introduction}

When a Josephson junction is fabricated from a semiconductor, its superconducting properties depend on the semiconductor's density of states.
This principle is the basis of the supercurrent transistor~\cite{Jarillo-Herrero2006a}, in which the junction's critical current is modulated by a nearby gate voltage, allowing the device to be switched between resistive and superconducting states.
Supercurrent transistors are components of low-temperature electronics such as SQUID magnetometers~\cite{Cleuziou2006} and gatemon qubits~\cite{Larsen2015, DeLange2015}.
They can be used to measure level crossings~\cite{Delagrange2016} and chiral states~\cite{Szombati2016} in junctions containing quantum dots, and also to investigate correlated-electron behaviour such as charge localisation~\cite{EstradaSaldana2019}.

Josephson junctions based on nanotubes and nanowires have been previously characterised in direct-current (DC) transport~\cite{Buitelaar2003, Jarillo-Herrero2006a, Cleuziou2006}.
Much greater sensitivity and speed can be achieved using the technique of radio-frequency (RF) reflectometry.
Furthermore, DC transport is sensitive only to the conductance of a junction but reflectometry is also sensitive to reactance, and therefore should enable measurements of quantum capacitance~\cite{Petersson2010} and Josephson inductance~\cite{Baumgartner2020}.
This has been confirmed by measuring the changes in conductance and inductance associated with the onset of superconductivity in NbTiN/InSb nanowire Josephson junction~\cite{Pei2014}.
The high-frequency impedance of nanowire Josephson junctions has also been investigated by integrating them into on-chip microwave cavities to measure the Andreev states~\cite{Hays2018, Tosi2019, Murani2019}.
However, the RF behaviour of a supercurrent transistor, in which the Josephson junction is switched on and off using a gate voltage, has never been investigated.

Here, we use simultaneous RF reflectometry and DC measurements to characterise a supercurrent transistor made of a carbon nanotube.
We compare these two methods across the full operating regime of the device in bias and gate voltage.
The RF data reproduces all the main features of the DC data, including the onset of superconductivity, a critical current that is tuned by a gate voltage and magnetic field, and the presence of Andreev reflections.
In addition, the noise is much lower, as expected for a high-frequency measurement.
These results show that a supercurrent transistor can be measured at RF without affecting its operation, and make RF techniques promising for rapidly characterising such devices under a range of conditions.

\section{Device and Measurement Setup}

\begin{figure}[t]
	\centering
	\includegraphics[width=0.95\columnwidth]{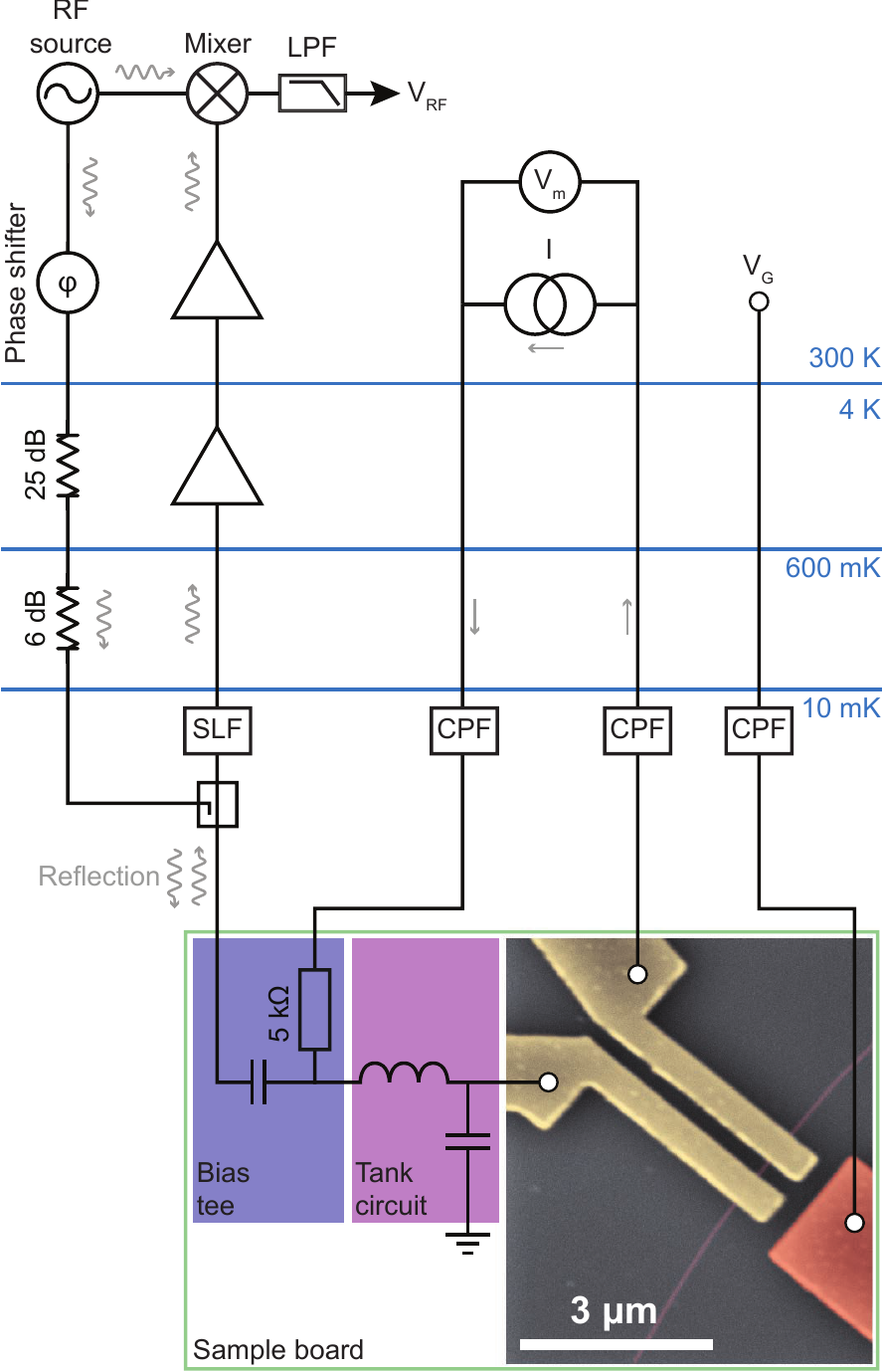}
	\caption{Measurement setup and device. The false-colour SEM image shows a carbon nanotube supercurrent transistor, with the nanotube highlighted in pink, the source and drain contacts in gold, and the gate electrode in red. The device is mounted on a printed circuit board containing a resonant tank circuit and a bias tee. This allows for simultaneous measurements by RF reflectometry, using the circuit on the left, and by current-biased DC transport, using the circuit in the centre. Wiggling arrows denote the RF path and straight arrows the DC current path. The gate voltage $V_\mathrm{G}$ is used for electrical tuning of the active transistor region. Here LPF denotes a low-pass filter, CPF a copper-powder filter containing an embedded low-pass $RC$ filter, and SLF a stripline filter consisting of a strip of copper embedded in eccosorb.}
	\label{fig1}
\end{figure}

Supercurrent transistors are fabricated using a single carbon nanotube contacted by superconducting electrodes as shown in figure~\ref{fig1}.
Device fabrication begins by growing nanotubes on a Si/SiO$_2$ substrate.
The nanotubes are grown by chemical vapour deposition using Fe/Ru catalyst nano particles~\cite{Li2007} at a temperature of 850~$^\circ$C with methane as the precursor gas~\cite{Samm2014}.
Bondpads and alignment markers are then patterned using electron beam lithography (EBL) and metalized with a bilayer of Ti/Au (10/50~nm).
Following lift-off, SEM imaging is used to locate and select individual nanotubes for the transistors.
Superconducting source and drain contacts and a gate electrode are then patterned with EBL, developed, cleaned in an ultraviolet ozone chamber, and metallised with a superconducting Pd/Al (4/80~nm) bilayer.
The edge-to-edge distance between the contacts is 300~nm.
Before mounting the sample, the room-temperature resistances of the devices are measured to check the fabrication yield.
Under atmospheric conditions and with the gate floating, roughly 80\% of the fabricated devices have a measurable conductance, with typical resistances between 7~k$\Omega$ and 100~k$\Omega$.
The two devices presented here (Devices A and B) are identical in design and fabrication procedure, but originate from separate fabrication batches, demonstrating the reproducibility of the fabrication.

The transistors are measured in the sample puck of a Triton dilution refrigerator with a base temperature between 10~mK and 20~mK, using the circuit shown in figure~\ref{fig1}, which allows simultaneous DC and RF measurements and in-situ impedance matching for optimal sensitivity~\cite{Ares2016b,Schupp2020}.
All DC wires are filtered using printed-circuit-board copper powder filters~\cite{Mueller2013} containing embedded RC filters.
The devices are biased with a source-drain current $\Ibias$, and gated using a voltage $\Vg$ applied to the side gate.
For DC resistance measurements, the voltage $\Vmeas$ at the top of the refrigerator is measured, and converted to a voltage $V$ across the device by subtracting the voltage drop across the inline series resistance using
\begin{equation}
V = \Vmeas-\Ibias \Rw.
\label{eq:V}
\end{equation}
Here $\Rw$ is the series resistance, which incorporates all ohmic resistances in the current path, and is determined by a linear fit to the $I-V$ trace in the supercurrent regime. It has the value $\Rw = 10.72~\mathrm{k}\Omega$~$(15.71~\mathrm{k}\Omega)$ for Device~A (Device~B), which is is consistent with the known inline resistance in the cryostat wiring.
The differential resistance $\partial V/\partial\Ibias$ is calculated numerically.

To allow for RF reflectometry measurements, the device is mounted on a sample board containing an $LC$ tank circuit and an in-built bias tee made from discrete chip components.
The loaded tank circuit exhibits a resonance frequency of 261.2~MHz with a quality factor of about 18.
Reflectometry measurements are performed by injecting an RF tone via attenuated coaxial lines and a directional coupler, reflected from the tank circuit, and fed into a homodyne demodulation circuit to generate a demodulated voltage $\VRF$, cf. figure~\ref{fig1}.
The RF power exciting the tank circuit was approximately $-117$~dBm.
Changes in resistance, capacitance, and inductance of the device modify the phase and amplitude of the reflected signal, and therefore the measured voltage $\VRF$~\cite{Roschier2004, Biercuk2006, Delbecq2011, Chorley2012, Ranjan2015}.

\section{DC spectroscopy}

\begin{figure*}[th]
	\centering
	\includegraphics[width=0.95\textwidth]{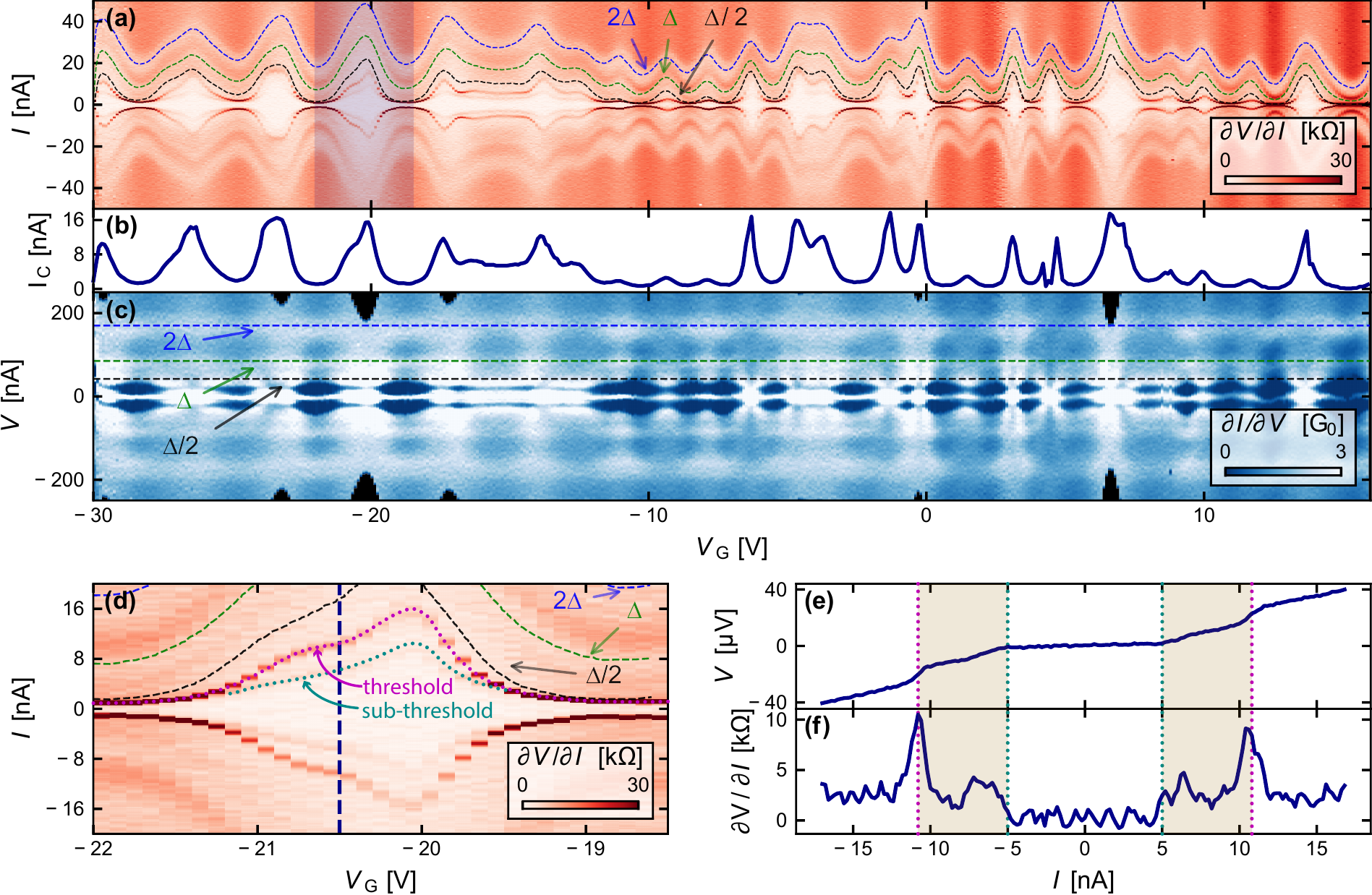}
	\caption{DC current- and voltage-bias spectroscopy, and sub-gap resistance of Device A, with the wiring resistance subtracted. (a) Differential resistance of the carbon nanotube as a function of the bias current $\Ibias$ and the gate voltage $\Vg$ for an interval of $\Vg$ in the p-type conduction regime. The resonances modulated with $\Vg$ are due to multiple Andreev reflections of different order. The first sharp resonance marks the critical current $\Ic$ of the Josephson junction. The white region near zero bias current, i.e. where $\partial V/\partial I=0$~k$\Omega$, is due to proximity-induced supercurrent flowing through the nanotube. The uniform region for large $\Ibias$ is due to quasiparticle transport. (b) Critical current $\Ic$ as a function of $\Vg$ extracted from (a). The maximum value measured in this device was $I_\mathrm{c,max}=17.7\pm0.7$~nA, where the error is the width of the transition in (a).
	(c) Differential conductance of the carbon nanotube, converted from (a), as a function of the source-drain voltage $V$ and the gate voltage $\Vg$ for the same interval as in (a).
	Dashed lines in (a), (c) and (d) mark expected positions of Andreev reflections, with $\Delta = 85~\mu$eV.
		(d) Zoom-in of differential resistance interval shaded blue in (a). 
		Two dotted lines mark the expected sharp threshold, from which $\Ic$ is taken, and the unexpected sub-threshold feature at $\Ibias<\Ic$.
		(e) I-V trace along dashed line in (d). (f) Differential resistance trace along dashed line in (d). (e) \& (f) The critical current is indicated by the purple dotted line. The shaded area between the purple and teal dotted lines indicates the region with a measurable sub-threshold resistance, whereas between the two teal dashed lines no residual resistance is observed.}
	\label{fig2}
\end{figure*}

To determine typical properties of the nanotube supercurrent transistor, we first characterized one device (Device A) using only DC measurements.
In this measurement, the current bias is applied directly to the bond pad of the source contact, bypassing the bias tee and tank circuit. Figure~\ref{fig2}(a) shows the differential resistance $\partial V / \partial \Ibias$ as a function of the bias current and gate voltage.
This measurement was performed in a gate voltage region more negative than the location of the band gap, i.e.\ in the $p$-type conduction regime.
As expected, we find the critical current is higher in this regime, because the Pd used as contacting layer in the superconducting leads yields a higher transparency of the contacts for holes~\cite{Chen2005}, i.e. larger $\Ic$.
When $\Ibias$ is close to zero, the differential resistance is small, indicating that the nanotube supports a supercurrent due to the proximity effect.
The boundary of this region is marked by a sharp peak in the differential resistance, and is taken as the superconducting critical current $\Ic$.

\begin{figure*}[th]
	\centering
	\includegraphics{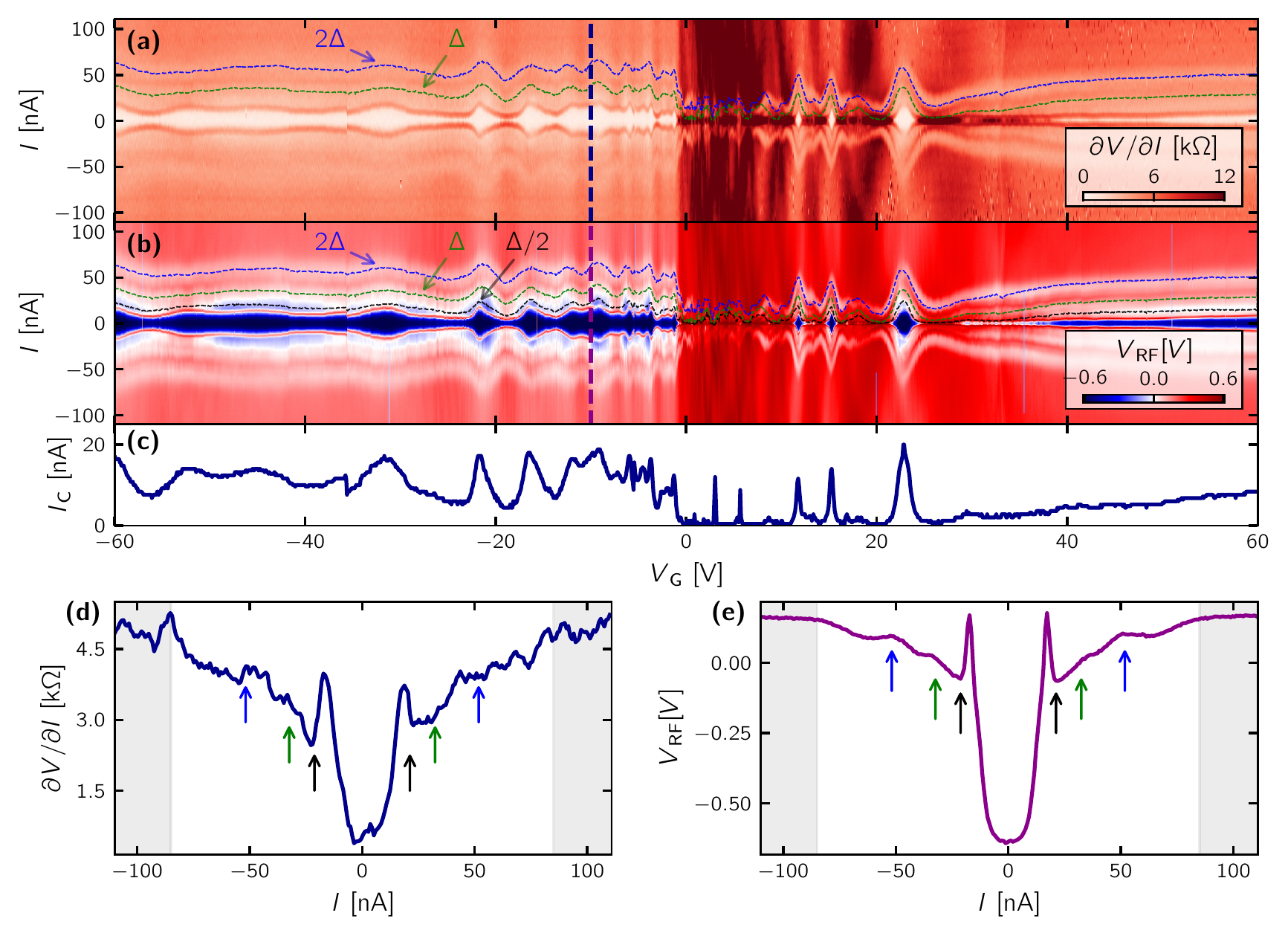}
	\caption{
	Current-biased measurements of a nanotube supercurrent transistor measured in DC transport (a) \& (d) and RF reflectometry (b) \& (e). 
	(a) Differential resistance as a function of $\Ibias$ and $\Vg$.
	(b) Simultaneously measured RF reflectometry signal. The main features observed in (a) also appear in the RF signal but with a higher contrast. 
	As in figure~\ref{fig2}, dashed lines in (a) and (b) mark expected positions of the Andreev reflections, using the measured $\Delta = 95~\mu$eV.
	(c) $\Ic$ as a function of $\Vg$, extracted from (b), with (a) giving similar results. In this device, $\Icmax=20 \pm 3$~nA.
	(d) Differential resistance for $\Vg=-10$~V (along the dashed line in (a). The critical current appears as two sharp peaks symmetrically placed around $\Ibias=0$~nA.
	Arrows mark the first three Andreev features.
	The region of quasiparticle transport (where the source-drain voltage exceeds $2\Delta$) is shaded grey. 
	(e) Equivalent trace plotting the RF voltage. The same features appear, but more clearly.
	}
	\label{fig3}
\end{figure*}

The critical current is strongly modified by changing $\Vg$, as seen from figure~\ref{fig2}(b).
It ranges from less than 0.1~nA 
to a maximum of $\Icmax=17.7$~nA.
These strong but irregular variations indicate corresponding fluctuations in the nanotube density of states, such as arise from Coulomb blockade or from Fabry-Perot oscillations~\cite{Jarillo-Herrero2006a}.
The maximum critical current is higher than in most previous experiments with nanotube junctions~\cite{Cleuziou2006,Schneider2012}, although a larger critical current of 30~nA has been achieved using niobium contacts~\cite{Pallecchi2008}.
This is a strong indication of high-transparency contacts between the superconducting leads and the nanotube, and demonstrates the quality of the contact provided by the Al/Pd bi-layer combined with a UV ozone cleaning prior to contact metal deposition. This progress in materials and fabrication methods allows for the integration of such nanotube Josephson junctions into more complex hybrid superconducting devices such as superconducting qubits~\cite{Mergenthaler2019}.
The large measured critical current may also indicate that this quantity depends on the number of walls of the nanotube. We did not measure the wall number or the nanotube diameter in our devices, and it is possible that the nanotube is multi-walled and that this allows for a larger criticial current.
The average critical current observed across the full region of $\Vg$ covered in figure~\ref{fig2}(b) is $\langle\Ic\rangle=4.7$~nA.
The fact that $\Ic$ depends on $\Vg$ confirms that the device is a supercurrent transistor.

For values of $\Ibias$ greater than $\Ic$, there is a series of broad dips in the differential resistance, consistent with multiple Andreev reflections~\cite{Nazarov2009, Buitelaar2003, Jarillo-Herrero2006a}.
To confirm this interpretation, figure~\ref{fig2}(c) shows the same data plotted against the source-drain voltage $V$ defined in Eq.~\eqref{eq:V}.
As expected, the peaks in conductance occur at voltages 
\begin{equation}
    V=\frac{2\Delta}{ne},
    \label{eq:andreev}
\end{equation}
where $n$ is an integer and $\Delta = 85~\mu$eV (Device A; 95$~\mu$eV, Device B), which is close to the typical value of the superconducting gap in nanotubes contacted with a thin film of aluminium~\cite{Buitelaar2003, Jarillo-Herrero2006a}.
The voltages calculated from equation~\eqref{eq:andreev} are marked by dashed lines in figure~\ref{fig2}(c) and align with features of low resistance as expected.

Inspection of figure~\ref{fig2}(a) also shows weak differential resistance even below the critical current threshold (i.e.\ for $|\Ibias| < \Ic$).
Figure~\ref{fig2}(d) is an expanded view showing this effect.
Two dotted lines highlight the resistance threshold at $I=\Ic$ and this weak sub-threshold feature.
Such a sub-threshold resistance indicates excitation away from the superconducting ground state.
Two possible causes are thermal phase diffusion and formation of a phase slip~\cite{Tinkham1996}.
Phase diffusion should lead to a smoothly increasing $\partial V/\partial\Ibias$~\cite{Ambegaokar1969}, and phase slips should lead to a series of abrupt steps in $V$~\cite{Tinkham1996}.
The sub-threshold feature in figure~\ref{fig2} does not follow either of these expectations. 
This is confirmed by figures~\ref{fig2}(e-f), which plot $V$ and $\partial V/\partial\Ibias$ along a single cross-section in bias current.
We therefore tentatively suggest that the sub-threshold peak in $\partial V/\partial\Ibias$ indicates that the device contains two weak links in series, and that the sub-threshold peak occurs when the weaker of the two becomes normal.
This might happen if the interface on one side of the device is less clean than on the other.
If this is the correct explanation, then the value of $\Ic$ plotted in figure~\ref{fig2}(c) is the critical current of the stronger link.
This sub-threshold peak is not observed in Device B (see below).
However, we note that a similar peak has been seen just below the transition temperature in a NbN nanowire device that also exhibited thermal and quantum phase slips~\cite{Masuda2016}.

\section{Spectroscopy using RF reflectometry}

Reflectometry experiments were performed on a second device (Device B) fabricated by a similar method as Device A but in a separate fabrication run.
Device B was bonded to the sample board in the same way as Device A (figure~\ref{fig1}) but now with the tank circuit connected.
The measurement of figure~\ref{fig2}(a) was now repeated, except that as well as the DC conductance $\partial V/ \partial I$, the demodulated RF voltage $\VRF$ was measured simultaneously.
The integration time per point was the same for the two data sets.

Figure~\ref{fig3}(a)-(c) shows the results.
As expected, the DC behaviour (figure~\ref{fig3}(a)) is similar to figure~\ref{fig2}(a), which confirms that the supercurrent transistor can operate with an RF excitation applied.
All the main features are reproduced in the RF measurement (figure~\ref{fig3}(a)), especially the sharp change at the critical current.
We find that $\VRF$ is approximately proportional to the DC resistance, showing that the RF measurement successfully transduces changes in the device impedance into changes in the tank circuit's reflectance.

Some features of the DC resistance appear more clearly in the RF data, in particular the sharp superconducting transition.
To explore this more thoroughly, figure~\ref{fig3}(d) and figure~\ref{fig3}(e) compare cross-sections at constant $\Vg=-10$~V, which is a typical gate voltage in the transistor's `On' configuration.
The superconducting transition and the first two Andreev features are evident in the conductance data (figure~\ref{fig3}(d)) but much clearer in the reflectometry data (figure~\ref{fig3}(e)).
The signal-to-noise ratio is higher for the RF measurement.
Quantifying the noise as the scatter of the data points in the quasiparticle transport regime, and the signal as the full vertical range of the traces in figure~\ref{fig3}, the RF measurement yields an improvement of $\sim 16$~dB.
This illustrates the advantage of measuring at an RF frequency where $1/f$ noise is reduced compared with DC.
It may also indicate that part of the RF signal is a response to the changing superconducting inductance, to which the DC measurement is insensitive.

\begin{figure}[t!]
	\centering
	\includegraphics[width=\columnwidth]{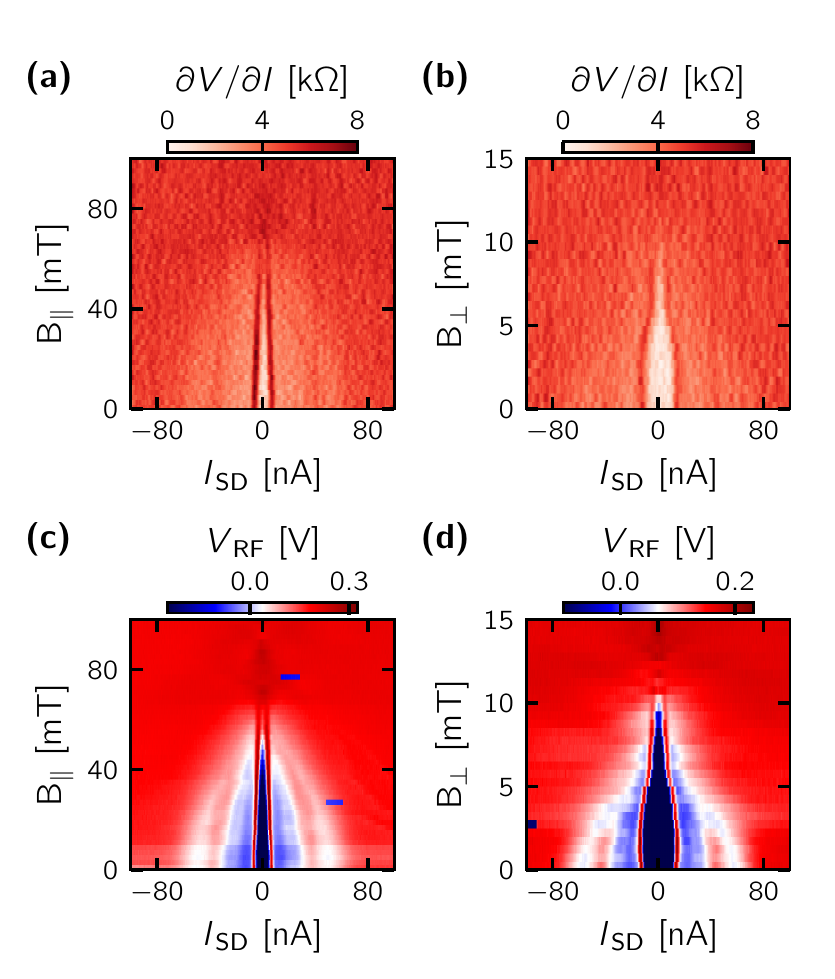}
	\caption{DC differential resistance (a-b) and RF signal (c-d) as a function of external magnetic field at $\Vg=-25$~V. The magnetic field is applied parallel (B$_\parallel$) as well as perpendicular to the chip surface (B$_\perp$).}
	\label{fig4}
\end{figure}

As a further illustration of the sensitivity of RF measurement, figure~\ref{fig4} shows the DC differential resistance and the RF signal as a function of magnetic field applied in (B$_\parallel$) and out of plane (B$_\perp$) of the device substrate.
As expected, the critical current decreases with increasing field, an effect that is seen more clearly in the RF than in the DC measurement.

\section{Conclusion}

By comparing simultaneous RF and DC transport measurements of a carbon nanotube supercurrent transistor, this experiment shows that RF reflectometry is sensitive to all the main transport features, most of which appear more distinctly than in DC transport alone.
Importantly, the properties of a supercurrent transistor device integrated into an RF measurement circuit are essentially identical to those of a device measured by DC transport alone.
Our results show that RF reflectometry is a non-invasive technique for characterising supercurrent transistors and potentially many other nanoscale devices and physical effects.
Although not tested here, reflectometry measurements can often be much faster than transport alone, and thus may allow many devices to be tested quickly under a wide range of operating conditions.
This would allow, for example, rapid fluctuations of the critical current to be measured in real time~\cite{Rogers1983}.
The reflectometry circuit used here might also allow for fast testing of other kinds of superconducting hybrid devices, for example to compare different ways of optimising superconductor-semiconductor interfaces, which are crucial for such devices.
The ability to distinguish reactive and resistive impedance changes is a possible tool for studying novel hybrid devices such as those used to realise Majorana qubits~\cite{Aguado2020, Smith2020}.

\acknowledgements{We acknowledge support from the Royal Academy of Engineering, EPSRC (EP/R029229/1), the ERC (818751), the Swiss Nanoscience Institute (SNI) and the Swiss National Science Foundation. M. M. acknowledges support from the Stiftung der Deutschen Wirtschaft (sdw). We thank Yu. A. Pashkin for discussions.}

\vfill


%

\end{document}